\documentclass[aps,pre,twocolumn,superscriptaddress]{revtex4-1}

\usepackage{graphicx,amsmath}
\usepackage{color,soul}
\usepackage{xcolor}
\newif\ifcolor

\newcommand{\cchange}[1]{%
\ifcolor
\textcolor{black}{#1}%
\else
#1%
\fi
}

\newcommand{\oran}[1]{%
\ifcolor
\textcolor{black}{#1}%
\else
#1%
\fi
}

\colortrue
\begin{document}

\title{Control of ecological outcomes through deliberate parameter
changes\\
in a model of the gut microbiome}


\author{Zipeng Wang}

\author{Eric W. Jones}
\email[]{ewj@physics.ucsb.edu}
\affiliation{Department of Physics, University of California, Santa Barbara,
    Santa Barbara, California, USA}

\author{Joshua M. Mueller}
\author{Jean M. Carlson}
\affiliation{Department of Physics, University of California, Santa Barbara,
    Santa Barbara, California, USA}
\affiliation{Interdepartmental Graduate Program in Dynamical Neuroscience, 
University of California, Santa Barbara, Santa Barbara, California, USA }


\date{\today}

\begin{abstract}

The generalized Lotka-Volterra (gLV) equations are a mathematical proxy for 
ecological dynamics.
We focus on a gLV model of the gut microbiome, in which
the evolution of the gut microbial state is determined in part by pairwise
inter-species interaction parameters that encode environmentally-mediated
resource competition between microbes.
We develop an \textit{in silico} method 
that controls the steady-state outcome of the system 
by adjusting these interaction parameters. 
This approach is confined to a bistable region of the gLV model.
In this method, a dimensionality reduction technique called steady-state reduction
(SSR) is first used to generate a two-dimensional (2D) gLV model that approximates the
high-dimensional dynamics on the 2D subspace spanned by the two steady states. 
Then a bifurcation analysis of the 2D model 
analytically determines parameter modifications that
    drive an initial condition to a \cchange{target} steady state.
This parameter modification of the reduced 2D model guides
parameter modifications of the original high-dimensional model, resulting in a 
change of steady-state outcome in the high-dimensional model.
This control method, called SPARC (\underline{S}SR-guided \underline{par}ameter 
\underline{c}hange), bypasses the
computational challenge of directly determining parameter modifications in the
original high-dimensional system.
SPARC could guide the development of indirect
bacteriotherapies, which seek to change microbial
compositions by deliberately modifying gut environmental variables such as 
gut acidity or macronutrient availability.
\end{abstract}


\maketitle

\section{\label{intro}Introduction}

A shared goal in environmental management, ecology, and medicine is to drive 
an ecosystem towards a target community structure.
For example, ocean and lake ecosystems benefit from the suppression of 
algal blooms, the control of invasive fish species helps preserve the
biodiversity of local fish populations, and certain microbial compositions
of the gut microbiome that
resist pathogenic infections improve the health of the host 
\cite{Algalbloom,Fish,Stein}. It is common to control these
ecosystems by directly altering the ecological composition of the community: 
unwanted algae can be removed by clay, invasive fish species can be killed by
biocides, and gut pathogens can be killed by antibiotics
\cite{Algalbloom,Thresher2014,antibiotic_treatment}.

In contrast to these direct methods that modify the ecological state of the
system, \textit{indirect} methods can control 
ecological outcomes by modifying environmental variables which effectively
change the dynamical landscape of the system
\cite{Jones}. 
For example, indirect control methods commonly applied to the previously
mentioned systems include reducing nutrient concentrations in water to inhibit 
algal blooms, lowering the water level to disrupt the spawning of invasive fish, 
and introducing prebiotics to promote biodiversity in the gut microbiome
\cite{Algalbloom,FishControl,prebiotics}.

In this paper, we create an \textit{in silico} technique 
that drives an ecological model towards a target outcome
by manipulating parameters that correspond to coarse-grained interactions between populations. 
Specifically, we seek a finite-time modification of the dynamical
landscape that drives an arbitrary initial condition towards a target state.
Although the intervention is temporary, the change in the ecological outcome
can be permanent. 

This control method is demonstrated in the context of \cchange{a data-driven
model} of the gut microbiome \cite{Buffie62}. 
Ecological dynamics are simulated using the generalized Lotka-Volterra (gLV) equations, a
commonly used model in theoretical ecology \cite{TAYLOR1988}.
In these equations, species-species interaction parameters represent
environmentally-mediated competition for resources.
These systems are often modeled by high-dimensional gLV equations in order to capture
the dynamics of the large number of microbial species that inhabit the gut
microbiome. In these models, the many inter-species feedbacks lead to complex dynamics. 
Accordingly, it is difficult to achieve a target steady-state outcome by
naively modifying parameter values, as such an approach requires exhaustively
searching a large parameter space.

To address these challenges, we focus on a bistable region of the ecological phase space 
that includes one \cchange{ target steady state and
another alternative steady state} within the gLV system. 
Then, a dimensionality-reduction technique called
steady-state reduction (SSR) is used to approximate the bistable region of interest
and to create a low-dimensional system with a compressed set of
interaction parameters \cite{EricSSR}. 
A bifurcation analysis of this 2D system determines 
a parameter modification that produces a targeted change in steady-state outcome.
Lastly, the low-dimensional interaction parameter change is associated with a
parameter change in the high-dimensional
model, which drives the original system to the target state.

This control method, referred to as SPARC (\underline{S}SR-guided
\underline{par}ameter \underline{c}hange), is applied to an 11-dimensional gLV model fit
to time-series data from a mouse microbiome experiment \cite{Buffie62,Stein}.
\cchange{ In this experimentally-derived gLV model, SPARC successfully alters
the steady-state outcomes of initial conditions by modifying interaction
parameters of the model.}
\cchange{SPARC as an \textit{in silico} approach is effective when applied to generic gLV
    systems, but its applicability to real-world systems is dependent on the fidelity of 
    the underlying gLV model.}
More generally, this method offers a systematic understanding of how environmental
factors and species-species interactions can be manipulated to control ecological
outcomes. 

\section{\label{method}Materials and methods}

\subsection{Generalized Lotka-Volterra equations}
The generalized Lotka-Volterra (gLV) equations are a traditional model in theoretical
ecology. Due to their flexibility, 
gLV models have been used to describe a wide variety of system dynamics, including
the market values of firms in the stock market, wolf
predation of multiple prey species, and the infection dynamics of RNA viruses
\cite{Market,Wolfs, fort2018predicting}.
In context of the microbiome, the gLV equations
have been used to model the population dynamics of gut microbial
communities \cite{venturelli2018deciphering}, and are given by

\begin{equation}
    \label{eq:high_D_gLV}
    \frac{d}{dt} y_i(t) = y_i(t)\Big( \rho_i  + \sum_{j=1}^N  K_{ij} y_j(t)\Big),
\end{equation}
where $y_{i}(t)$ denotes the abundance of microbes of species $i$ at a given 
time $t$, $\rho_{i}$ is the growth rate of species $i$, and $K_{ij}$ is the 
interaction coefficient between two populations $i$ and $j$. 
The interaction parameters $K_{ij}$ form the $N\times N$ interaction matrix
$K$, where $N$ is the number of species.
The growth rate parameters $\rho_i$ are constrained by $\rho_i>0$. 
The interaction parameters $K_{ij}$ capture prototypical
ecological interactions such as competition, symbiosis, and amensalism
\cite{venturelli2018deciphering}.
Specifically, the parameter $K_{ij}$
represents the effect of species $j$ on species $i$, which is mediated
by environmental factors such as available nutrients. 
Thus, if environmental factors are changed,
the parameters $K_{ij}$ will change as well. 

\cchange{In general, gLV systems can exhibit periodic and
chaotic behaviors \cite{FAN199947,vano_2006}, and the criteria that predict the
stability of ecosystems based on their structure have been prominently studied
in theoretical ecology \cite{MAY1972,Allesina2012,Gibbs}. 
Here, we focus on ecological dynamics that relax to point attractors. In this
regime, the gLV dynamics of concern can be represented by a pseudo-energy landscape
(e.g. a Lyapunov function), which is a scalar field in ecological state space
that behaves analogously to a physical energy landscape.}

\begin{figure}
    \includegraphics[width=1\columnwidth]{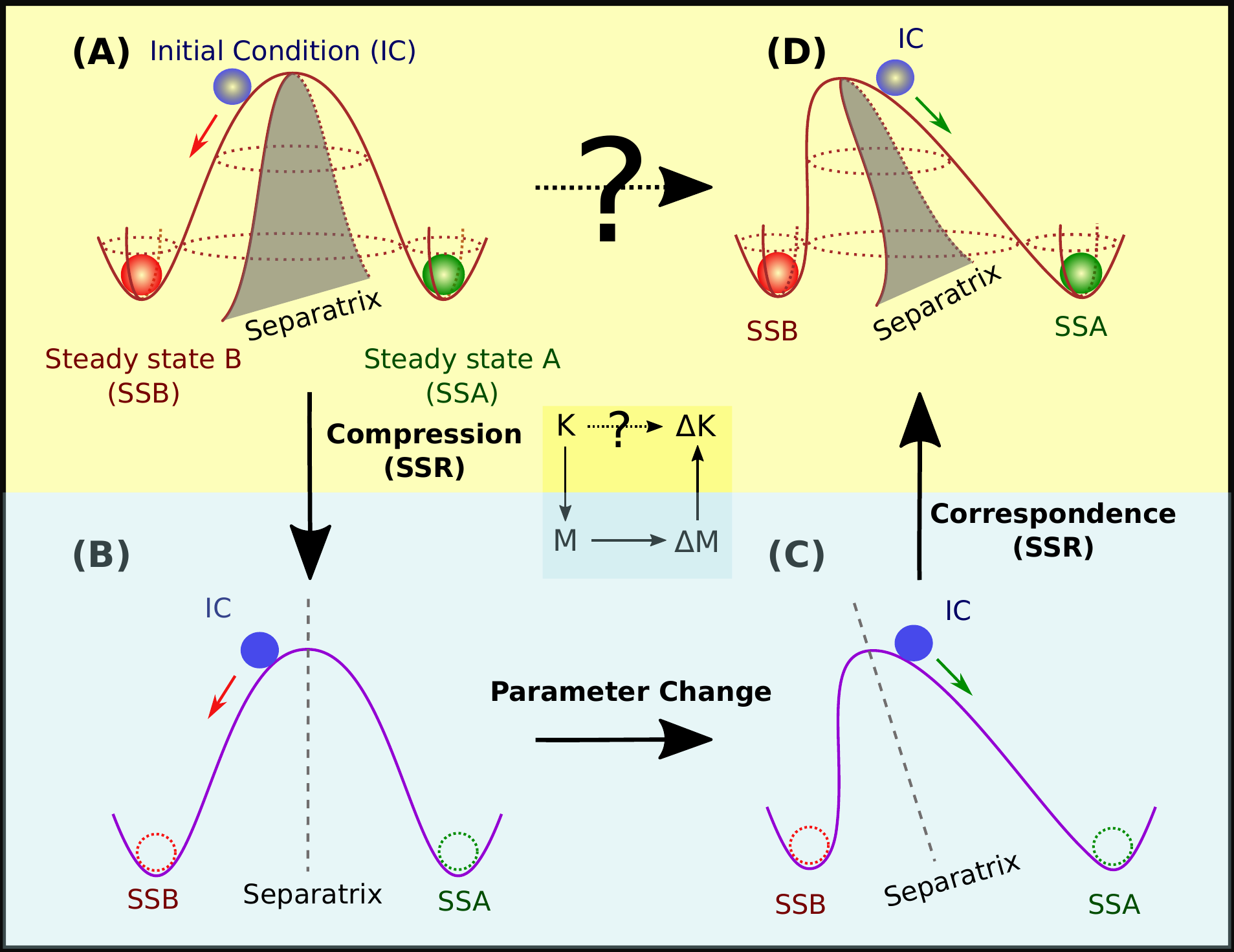}
    \caption{\label{fig:schematic}
    \textbf{A schematic overview of how SPARC (SSR-guided parameter change)
     controls steady-state outcomes.}
    (A) A bistable region in a high-dimensional gLV model, with two steady
    states and an initial condition tending towards the \cchange{alternative} steady state
    (shown in red), is represented as a pseudo-energy (Lyapunov) landscape. This
    landscape is parameterized by the interaction matrix $K$ of the
    high-dimensional gLV system.
    (B) The high-dimensional landscape is compressed into a reduced
    2-dimensional landscape, generated by the dimensionality-reduction technique 
    steady-state reduction (SSR) as described in Eq.~(\ref{SSR}). This 2D landscape is
    parameterized by a $2\times 2$ interaction matrix $M$. 
    (C) Guided by a bifurcation analysis of this reduced 2D system, a
    modification of the interaction matrix $\Delta M$ changes the Lyapunov
    landscape in a targeted way. 
    After this change, the initial condition tends 
    towards the healthy steady state (shown in green) in the low-dimensional
    system.
    (D) A high-dimensional parameter modification $\Delta K$, informed by the
    2D parameter change $\Delta M$ via the SSR formulae,
    changes the high-dimensional Lyapunov landscape. 
    It is computationally difficult to identify this parameter change directly from the
    original model (A to D), but using SSR and the bifurcation analysis of the
    2D model, this change is straightforward (A to B to C to D). 
    }
\end{figure}

We wish to determine a coordinated modification of these interaction
parameters $K_{ij}$ that drive the system to a target state.
The growth rates $\rho_i$ and interactions $K_{ij}$ determine the
dynamical landscape on which the microbial system evolves. A
modification of the interaction parameter matrix $K$
reshapes the dynamical landscape of the gLV system. 
This reshaping process is visualized schematically in
Fig.~\ref{fig:schematic}. 

All simulations in this paper were run with the quadrature method
\texttt{odeint} from the Python module \texttt{scipy.integrate}.

\subsection{A gLV model fit to experimental data}

\cchange{In a mouse experiment, Buffie \textit{et al.}~demonstrated that mice that are
administered the antibiotic clindamycin become 
susceptible to \textit{Clostridiodes difficile} infection (CDI)
\cite{Buffie62}. }
Stein \textit{et al.}~fit
a gLV model, referred to as the CDI model,
to the time-series microbial abundance data \cchange{from} this mouse experiment 
\cite{Stein}. 
\cchange{For modeling purposes, microbial species are coarse-grained at the genus level,
resulting in 11 microbial populations, each described by a population $y_i$ in
the gLV model.}
This gLV model
captures the CDI-resistant and CDI-susceptible steady states that are observed
in the experiment \cite{Stein,Jones}.

\cchange{
The dynamical structure of this CDI model is characterized by the composition and
stability of its steady states. Two steady states in this model, including the
experimentally-observed CDI-resistant state, are locally stable (i.e., the
eigenvalues of the Jacobian matrix evaluated at these steady state compositions
are exclusively negative). Additionally, the CDI model features six steady
states whose Jacobian matrices have one non-negative eigenvalue (referred to as
having one ``unstable direction"); it also features 23 steady states, including the
CDI-susceptible state, whose Jacobian matrices have two non-negative
eigenvalues (i.e., with two unstable directions). The CDI-susceptible state is
composed of 5 coarse-grained species and the CDI-resistant state is composed
of 3
coarse-grained species. In these steady states, the abundance of all other
species is zero. The detailed compositions of these steady states are given in
the Supplementary Information.
}

\cchange{
In this paper, we examine the transition between the CDI-susceptible state and
the CDI-resistant state, and apply SPARC to the bistable region formed by these
states.} \oran{First, we demonstrate SPARC in the ``infection" scenario in which
the CDI-susceptible steady state is 
treated as the target state and the CDI-resistant steady state is designated the
alternative state.}
\cchange{We consider an initial condition on
the plane spanned by the target state $\vec y_a$ and the alternative state $\vec y_b$  that tends
towards $\vec y_b$ in the absence of any intervention. The goal of SPARC is to find
a modification to the interaction matrix 
$\Delta K$ that alters the evolution of this initial condition
and drives it towards the target state.} \oran{In the infection scenario, this parameter change represents a
disruption of the microbial dynamics that can drive the system towards a state
susceptible to CDI.}

\cchange{We later consider the} 
\oran{``recovery" scenario} \cchange{in which the target state 
is the CDI-resistant state and the
alternative state is the CDI-susceptible state. In this scenario too, SPARC
alters the steady-state behavior of an initial condition so that it flows
towards the target CDI-resistant state.}
\oran{The parameter change generated by SPARC in this scenario informs the
intervention needed to recover from the CDI-susceptible state in this model.}
\cchange{These results demonstrate that, 
for this pair of steady states in the CDI model, SPARC
is able to drive microbial dynamics in the direction of either steady state.}\\

\subsection{SSR-guided parameter change (SPARC)}

We develop a multi-step control
framework to determine a parameter change that drives a given initial
condition towards a target state. A bistable
landscape of interest in a high-dimensional gLV model is first reduced into a
2D gLV model using steady-state reduction (SSR) \cite{EricSSR}.
This control framework is called SPARC (\underline{S}SR-guided 
\underline{par}ameter \underline{c}hange), and summarized
in Fig.~\ref{fig:schematic}.

\subsubsection{Steady-state reduction}

Steady-state reduction (SSR), developed by Jones and Carlson, 
is a mathematical technique that compresses a high-dimensional gLV system
into a 2D gLV system, as shown in Fig.~\ref{fig:schematic}A and
B \cite{EricSSR}.
In a high-dimensional gLV model of $N$ species, there are 
$N^2$ interaction parameters. Due to the complexity of the feedbacks of the
ecological system, 
it is analytically intractable and computationally expensive to numerically determine
how modifications of interaction parameters affect the asymptotic behavior of arbitrary 
initial conditions. 

To understand the dynamics in the high-dimensional phase
space, we consider bistable systems and focus on the subspace
spanned by the two steady states $\vec{y}_a$ and
$\vec{y}_b$. The SSR technique views steady states $\vec{y}_a$ and
$\vec{y}_b$ of the high-dimensional model as idealized composite states and
constructs a new set of 2D gLV equations in which the basis vectors
correspond to the
high-dimensional steady states. This 2D gLV system approximates
the slow manifold that connects $\vec{y}_a$ and $\vec{y}_b$, and is the
\cchange{best} 
possible \cchange{gLV} approximation of the high-dimensional dynamics on the subspace spanned by
$\vec{y}_a$ and $\vec{y}_b$ \cite{EricSSR}. Explicitly, the
approximate 2D gLV system has the form 

\begin{align}
    \frac{dx_a}{dt} &= x_a(\mu_a+M_{aa}x_a+M_{ab}x_b), \nonumber  \text{ and} \\
    \frac{dx_b}{dt} &= x_b(\mu_b+M_{ba}x_a+M_{bb}x_b),  \label{2DgLV}
\end{align}
where $x_a$ corresponds to the high-dimensional gLV system's component in the 
direction $\hat{x}_a = \frac{\vec{y}_a}{\left\lVert\vec{y}_a\right\rVert}_2$,
$x_b$ corresponds to the direction $\hat{x}_b =
\frac{\vec{y}_b}{\left\lVert\vec{y}_b\right\rVert}_2$, 
and $\left\lVert\vec{v}\right\rVert_2$ is the 2-norm of $\vec{v}$. The
parameters $\mu_a$ and $\mu_b$ represent the growth rates of $x_a$ and $x_b$,
and the $M_{ij}$ interaction parameters form a 2D interaction matrix $M$. 
\cchange{SSR yields the reduced interspecies interaction parameters $M_{ab}$
and $M_{ba}$, which are given by}

\begin{widetext}
\begin{align}
    \label{completeSSR}
    M_{ab} &=
    \frac{\sum_{i,j=1}^NK_{ij}(y_{ai}y_{bj}+y_{bi}y_{aj})
    \left(y_{ai}-y_{bi}\sum_{k=1}^{N} y_{ak}y_{bk}\right)}
    {1-(\sum_{i=1}^Ny_{ai}y_{bi})^2},
    \nonumber \text{ and} \\
    M_{ba} &=
    \frac{\sum_{i,j=1}^NK_{ij}(y_{bi}y_{aj}+y_{ai}y_{bj})
    \left(y_{bi}-y_{ai} \sum_{k=1}^{N}y_{bk}y_{ak}\right)}
    {1-(\sum_{i=1}^Ny_{bi}y_{ai})^2},
\end{align}
\end{widetext}
\cchange{where $y_{ai}$ and $y_{bi}$ are the $i$th components of the unit
vectors $\hat y_a \equiv \vec y_a/ ||\vec y_a||_2$ and $\hat y_b \equiv \vec
y_b/ ||\vec y_b||_2$, respectively. The other 2D parameters $\mu_a$, $\mu_b$, $M_{aa}$, and $M_{ab}$ are given by}

\begin{eqnarray}
    \mu_\gamma &= \frac{\vec{\rho}\cdot \vec{y}_\gamma^{\circ 2}}{\left
    \lVert\vec{y}_\gamma\right\rVert^2_2}, \text{ and} \nonumber \\
    M_{\gamma\delta} &= \frac{(\vec{y}_\gamma^{\circ 2})^TK\vec{y}_\delta}
    {\left\lVert\vec{y}_\gamma\right\rVert^2_2\left\lVert\vec{y}_\delta
    \right\rVert_2} \label{SSR},
\end{eqnarray}
\cchange{where $\gamma,\delta \in a,b$. When the high-dimensional steady states $\vec
y_a$ and $\vec y_b$ are orthogonal, the interspecies interaction parameters
$M_{ab}$ and $M_{ba}$ in Eq.~(\ref{completeSSR}) reduce to the interaction
parameters in Eq.~(\ref{SSR}). In these formulae, }
$\vec{y}^{\circ 2} \equiv \text{diag}(\vec{y})\vec{y}$ is the 
element-wise square of $\vec{y}$.
Note that SSR maps the high-dimensional steady states $\vec y_a$ and $\vec y_b$
to the points \oran{$(||\vec y_a||_2,0)$ and $(0,||\vec y_b||_2)$},
which are the steady states of the 2D model.
Additionally, if the high-dimensional steady states are stable, SSR guarantees
that their low-dimensional counterparts are stable as well.
The fidelity of the SSR method is demonstrated in Fig.~\ref{fig:Result_4_panel},
where it is applied to an experimentally-derived gLV system. Additional
examples are provided in the Supplementary Information.

\subsubsection{Bifurcation analysis}

After the high-dimensional gLV model is reduced to a 2D model, the next step is
to find a parameter change in the 2D model that changes the steady-state
behavior of the system, as shown in Fig.~\ref{fig:schematic}B and C.
Simplifying the high-dimensional system 
using SSR results in a 2D gLV model with two growth rate parameters, $\mu_a$ and
$\mu_b$ and four interaction parameters, 
$M_{aa}$, $M_{ab}$, $M_{ba}$, and $M_{bb}$.

When the steady states of the original high-dimensional bistable system are
stable, SSR guarantees two stable steady states at $(1,0)$ and $(0,1)$. 
In addition to these two steady states, the system possesses a trivial unstable
steady state at $(0,0)$, and another
hyperbolic fixed point with nonzero $x_a$ and $x_b$ components. The separatrix,
which delineates the basins of attraction of the $(1,0)$ and $(0,1)$ steady
states, is topologically required to pass through this hyperbolic fixed point.

\begin{figure}
\includegraphics[width=1\columnwidth]{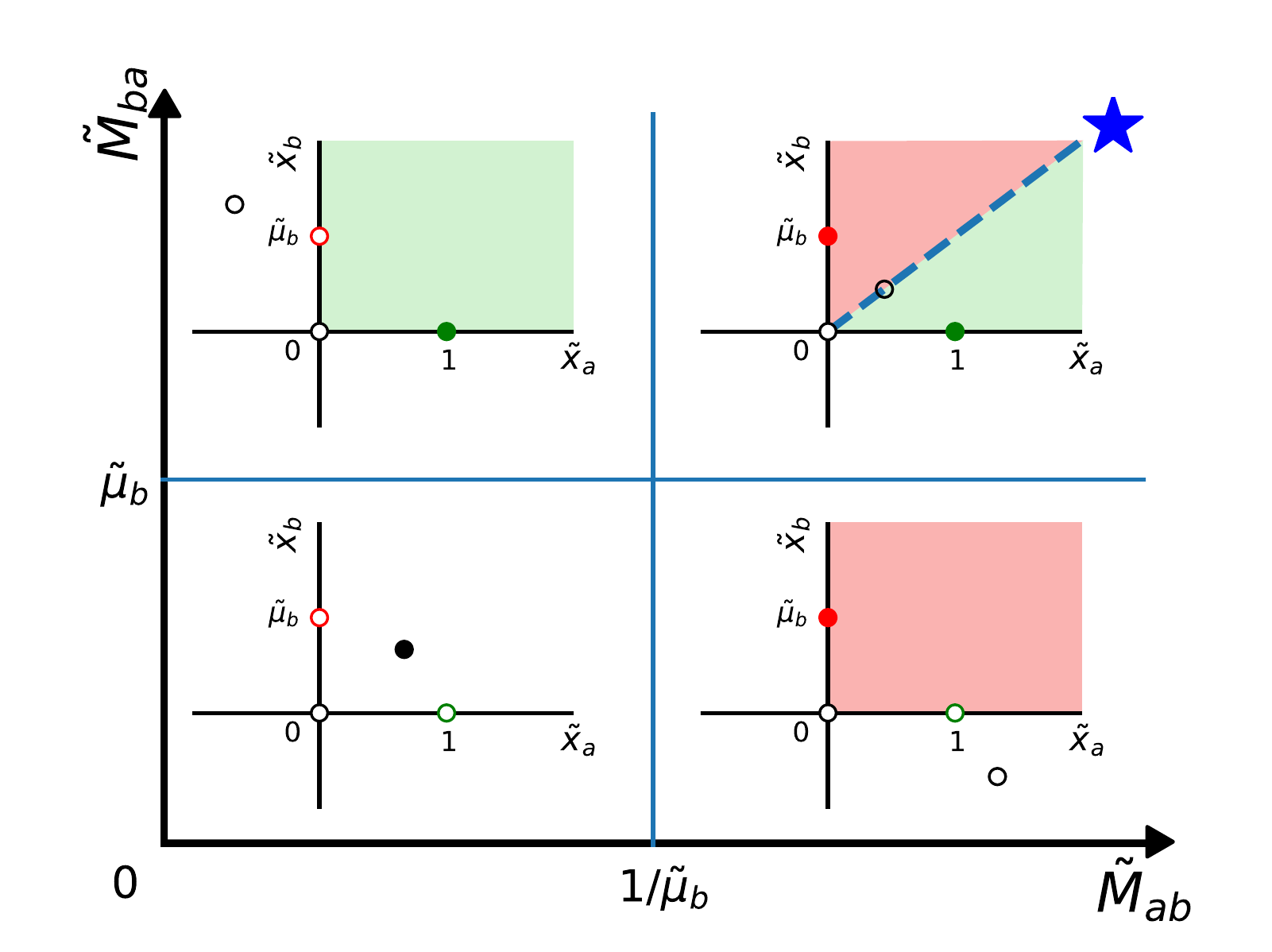}
    \caption{\label{fig:bifur}
    \textbf{A bifurcation diagram of nondimensionalized
2D gLV systems.}
    This diagram shows phase space representations of different topological
    classes of 2D gLV dynamical landscapes,  
    and their dependence on the nondimensionalized parameter values $\tilde
    M_{ab}$, $\tilde M_{ba}$, and $\tilde \mu_b$ of Eq.~(\ref{nondim}). 
    The lines at $\tilde
    M_{ab} = 1/\tilde \mu_b$ and $\tilde M_{ba} = \tilde \mu_b$ split the
    parameter space into four quadrants that each correspond to a different 
    topological configuration of phase space.
    The graph inside each quadrant shows a representative phase space
    configuration of the
    nondimensionalized gLV system, where $\tilde x_a$ and $\tilde x_b$ are the
    rescaled populations in Eq.~(\ref{nondim}).
    The hollow dots represent unstable steady
    states, and the filled dots represent stable steady states. 
    The basins of attraction of the steady states 
    $(1,0)$ and $(0,\tilde \mu_b)$ are shaded in green and red, respectively.
    The upper-right quadrant, labeled with a blue star, represents
    the parameter regime in which bistable 2D landscapes occur. 
    An alternative visualization of this bistable landscape is schematized in
    Fig.~\ref{fig:schematic} as a pseudo-energy landscape. The reduced 2D
    gLV models, generated by applying SSR to bistable regions in
    high-dimensional gLV models, reside in this upper-right quadrant. 
    In this bistable quadrant, 
    the separatrix passes through the hyperbolic steady state with
    non-negative coordinates. The steady states at $(1,0)$ and $(0,\tilde \mu_b)$ 
    undergo transcritical bifurcations in response to changes in 
    $\tilde M_{ab}$ and $\tilde M_{ba}$,
    yielding the diagrams in adjacent panels. The lower-left quadrant is included for completeness.}
\end{figure}

When nondimensionalized, the 2D gLV equations Eq.~(\ref{2DgLV}) become
\begin{align}
    \frac{d\tilde{x}_a}{dT} &=
    \tilde{x}_a(1-\tilde{x}_a-\tilde{M}_{ab}\tilde{x}_b),
    \nonumber \text{ and} \\
    \frac{d\tilde{x}_b}{dT} &=
    \tilde{x}_b(\tilde{\mu}_b-\tilde{M}_{ba}\tilde{x}_a-\tilde{x}_b),
    \label{nondim}
\end{align}
where $\tilde{x}_a=-\frac{M_{aa}}{\mu_a}x_a$, 
$\tilde{x}_b=-\frac{M_{bb}}{\mu_a}x_b$, $T = \mu_at$, $\tilde{M}_{ab} =
M_{ab}/M_{bb}$, $\tilde{M}_{ba} = M_{ba}/M_{aa}$, and $\tilde{\mu}_b =
\mu_b/\mu_a$. In terms of these nondimensionalized parameters, the two steady
states are now at $(1,0)$ and $(0,\tilde{\mu}_b)$. The coordinate of the 
hyperbolic fixed point is given by

\begin{equation}
    \label{third_state}
    \left(\frac{\tilde{M}_{ab}\tilde\mu_b-1}{\tilde{M}_{ab}\tilde{M}_{ba}-1},
    \frac{\tilde{M}_{ba}-\tilde \mu_b}{\tilde{M}_{ab}\tilde{M}_{ba}-1}\right).
\end{equation}
Since the separatrix passes through this steady state,
adjusting the parameters $\tilde{M}_{ab}$ and
$\tilde{M}_{ba}$ alters its position and stability, as shown in Fig.~\ref{fig:bifur}. 

\cchange{A necessary condition for the steady states $(1,0)$ and $(0,\tilde \mu_b)$ to be stable
is that $\tilde M_{ab} \tilde M_{ba}-1>0$ \cite{EricSSR}. Thus,
when} $\tilde M_{ab}$ is made smaller than $1/\tilde \mu_b$ with $\tilde M_{ba}$ fixed,
the $\tilde x_a$ coordinate of the unstable steady
state becomes negative. Equivalently, in Fig.~\ref{fig:bifur} this corresponds
to system moving from the top-right configuration to the top-left
configuration. A linearized stability analysis finds that the topological
structure of the 2D phase space also changes after this parameter change is
made. As shown in the top-left panel of Fig.~\ref{fig:bifur}, the steady state
at $(0,\tilde \mu_b)$ becomes unstable once $\tilde M_{ab}$ is smaller than
$1/\tilde \mu_b$, which forces initial conditions in the top-right quadrant of the phase space
towards the stable state at $(1,0)$. Similarly, once
$\tilde M_{ba}$ is smaller than $\tilde \mu_b$, the $\tilde x_b$ coordinate of the
hyperbolic steady state becomes negative. In Fig.~\ref{fig:bifur} this
corresponds to crossing from the top-right to the bottom-right, at which point the steady state at
$(1,0)$ becomes unstable. The bifurcation diagram in Fig.~\ref{fig:bifur} 
provides a guide for how the steady-state structure of the 
2D gLV equations depends on the interaction parameters.  

This bifurcation analysis indicates how to move the separatrix in a particular direction. 
Numerical methods determine the minimal change
of parameters $\tilde{M}_{ab}$ or $\tilde{M}_{ba}$ that switch the
asymptotic steady-state 
behavior of a given initial condition.
In simulations where the target steady state is located at $(1,0)$, 
the value of $\tilde M_{ab}$ is decreased incrementally,
spanning from its original value through $1/\tilde \mu_b$. 
In terms of the dimensionalized 2D gLV system, this corresponds to keeping $M_{bb}$ 
constant while $M_{ab}$ is
modified until the separatrix is shifted to a position where the initial
condition switches from one basin of attraction to the other.

\subsubsection{Correspondence between 2D and high-dimensional gLV models}

Changes in the 2D interaction parameters that drive an initial condition to
a target state are associated with changes in the high-dimensional
interaction parameters, since the 2D reduced parameters are functions of the
high-dimensional parameters via the SSR formulae. This is schematically
shown in the transition from Fig.~\ref{fig:schematic}C to
Fig.~\ref{fig:schematic}D. More explicitly, Eq.~(\ref{SSR}) can
be re-written as 

\begin{equation}
    \label{2Dto11D}
    M_{\gamma\delta} = \sum_{i,j}\alpha_{ij}^{\gamma \delta}(\vec y_a, \vec y_b)K_{ij},
\end{equation}
where $\gamma,\delta \in \{a,b\} $, and $\vec y_a$ and $\vec y_b$ are the two
steady states of interest. In this paper, since the target state is placed at
$(1,0)$, it is most important to modify the parameter $\tilde
M_{ab} = M_{ab}/M_{bb}$. For simplicity we only consider modifications to
$M_{ab}$, and therefore are primarily concerned with the coefficients
$\alpha_{ij}^{ab}$, hereafter referred to as $\alpha_{ij}$. Thus, from this correspondence 
a modification in the 2D interaction matrix $M$  may be reproduced in the high-dimensional 
system by modifying the high-dimensional interaction matrix $K$.
This choice is degenerate --- there is more than one way to change the
high-dimensional interaction matrix $K$ that corresponds to the same 2D
parameter modification.  Note that the smallest possible
high-dimensional parameter change $\Delta K_{ij}$ is associated
with the largest coefficient $\alpha_{ij}$.

\section{\label{result}Results}

SPARC (SSR-guided parameter change) controls the steady-state outcome of a
high-dimensional gLV system by deliberately changing the geometry of its dynamical landscape.
~SPARC (i) approximates a bistable landscape of a high-dimensional 
gLV system by its 2D SSR-generated counterpart, 
(ii) identifies a 2D interaction parameter change that switches the asymptotic
behavior of an initial condition on this bistable landscape, and (iii)
associates the 2D
parameter change with a parameter modification in the high-dimensional gLV system. This
parameter modification shifts the high-dimensional landscape so that
an otherwise disease-prone initial condition will instead tend towards the
target state. 

\begin{figure}
\includegraphics[width=1\columnwidth]{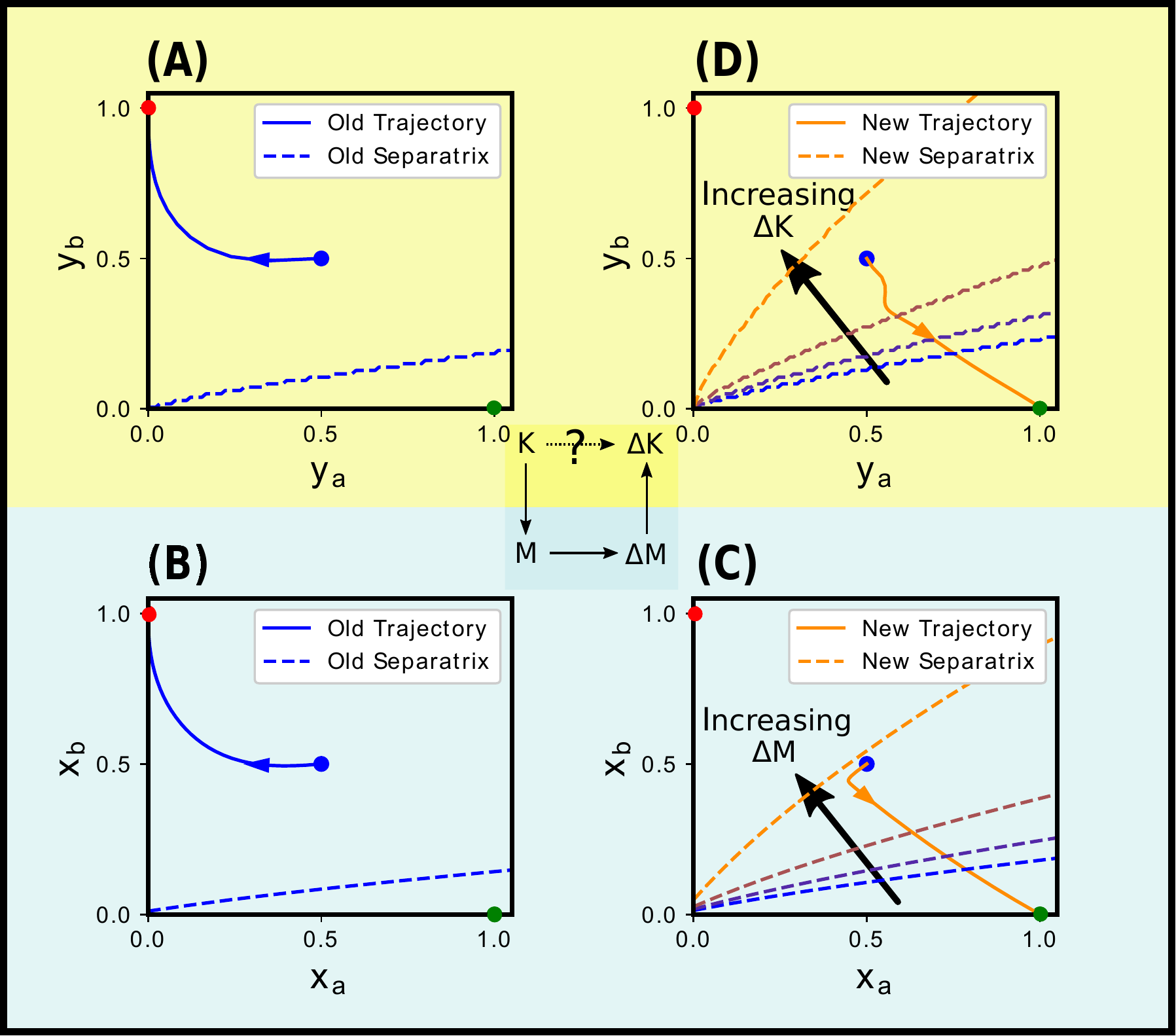}
 \caption{\label{fig:Result_4_panel}
    \textbf{A realization of SPARC, 
    as described in Fig. 1, applied to the infection scenario of the CDI model.}
    (A) The phase space of the CDI model~\cite{Stein} 
    is projected onto the
    2D plane spanned by the \cchange{target} steady state $\vec y_a$ and the
    \cchange{alternative} steady state $\vec y_b$. 
    \oran{The target and alternative steady states at  $(||\vec y_a||_2,0)$ and 
    $(0,||\vec y_b||_2)$ are rescaled in this plot to the points
    $(1,0)$ and $(0,1)$.}
    The in-plane separatrix, generated numerically, delineates the basins of 
    attraction. 
    (B) Steady-state reduction (SSR) generates an approximate 2D phase space. 
    Notice that the 2D separatrix and trajectory qualitatively resemble those in (A). 
    (C) The 2D separatrix moves as the 2D interaction
    matrix $M$ is modified. Four separatrices corresponding to four
    changes with increasing magnitude in the interaction matrix $\Delta M$ are
    shown.
    The matrix element and direction of this change are guided by the bifurcation
    analysis in Fig.~\ref{fig:bifur}.
    A sufficiently large parameter change alters the steady-state outcome
    of the initial condition $(0.5,0.5)$.
    (D) Changes in the low-dimensional interaction parameter $\Delta M$ are
    associated with changes in high-dimensional parameter $\Delta K$ by SSR
    formulae. The
    resulting shift in the high-dimensional separatrix is qualitatively similar
    to that of the low-dimensional system. In particular, the initial condition
    $(0.5\vec y_a + 0.5\vec y_b)$ now evolves towards the \cchange{target} steady state.
    SPARC successfully alters the steady-state outcome 
    without having to search a 121-dimensional parameter space.}
\end{figure}

Note that since the steady states of the high-dimensional model are dependent
on the interaction matrix $K$, a small change in this matrix will slightly modify
the coordinates of the steady states. Thus, to allow the system to evolve back to
the original steady states, this parameter modification must
be turned off after some time. \cchange{To initially demonstrate SPARC,
the parameter modification is turned off once the system stabilizes at the shifted
steady state (Fig.~\ref{fig:Result_4_panel}). When SPARC is applied to the CDI model,
the parameter modification is small enough that the changes in steady state
locations are negligible (Supplementary Information).
Later, when considering the ``recovery" scenario, the 
parameter modification is turned off before the system stabilizes at any steady
state; in this case there is a critical duration that
the parameter modification must be active for in order for the intervention to
be successful (Fig.~\ref{fig:correct}).} 

In this section, SPARC is first applied to the CDI
model fit by Stein \textit{et al.}~to data from a 
\textit{Clostridioides difficile} infection (CDI) experiment in mice
\cite{Stein}. Then, the robustness of SPARC is examined by applying it to synthetic gLV models.

\setlength\belowcaptionskip{-2ex}

\subsection{Steady-state reduction (SSR) produces a 2D
approximation to bistable dynamics in a high-dimensional gut microbiome model}

First, bistable dynamics in the CDI model are approximated 
by reduced dynamics on a 2D subspace generated by
steady-state reduction (SSR). 
We focus on two steady states of this gLV model that correspond to
experimentally observed 
CDI-resistant and CDI-susceptible microbiome compositions. \cchange{For the
initial demonstration of SPARC,} \oran{we consider the ``infection" scenario in
which} \cchange{ the CDI-susceptible state is defined 
as the target state and the CDI-resistant state is defined as the alternative
state.}

\cchange{The target state and the alternative state are represented by the
high-dimensional vectors $\vec{y}_a$ and $\vec{y}_b$, respectively.}
The microbial dynamics that result from the initial
condition $(0.5\vec{y}_a+0.5\vec{y}_b)$  tend towards the alternative steady
state $\vec y_b$. To visualize these dynamics, the trajectory is projected onto
a plane spanned by the steady states $\vec y_a$ and $\vec y_b$, as displayed in
Fig.~\ref{fig:Result_4_panel}A. In this figure, the axes are rescaled so that
the steady state $\vec y_a$ is located at point $(1,0)$ and the steady state $\vec y_b$ 
is located at point $(0,1)$.
The separatrix shown in 
Fig.~\ref{fig:Result_4_panel}A is numerically generated from trajectory
simulations. Notice that on this subspace, the initial condition is above the
separatrix, and hence the initial condition evolves towards the
\cchange{alternative} steady
state at $(0,1)$.

This 11D bistable landscape is approximated by a reduced 2D gLV model generated by SSR,
according to Eq.~(\ref{SSR}). The SSR-generated parameter values and their
nondimensionalized counterparts are provided in the Supplementary Information.
The dynamics of the reduced 2D trajectory were initial condition $(0.5,0.5)$
are displayed in 
Fig.~\ref{fig:Result_4_panel}B, and are similar to the projection of the 11D
dynamics in Fig.~\ref{fig:Result_4_panel}A. 
Note that the position of the separatrix, which is generated analytically in
the 2D model \cite{EricSSR}, is well-approximated by SSR.
In the Supplementary Information it is further demonstrated that this 
reduced 2D model accurately approximates the high-dimensional trajectories that originate 
from other initial conditions.

\cchange{
It is difficult to identify the interspecies feedbacks that induce
bistability in a high-dimensional system: in general, it is unclear how the
separatrix changes as a function of the system parameters.
On the other hand, in the reduced 2D gLV system, there
are well-defined conditions for 
bistability, namely
\begin{align}
    \tilde M_{ab} \tilde \mu_b &= (M_{ab}/M_{bb})(\mu_b /
    \mu_a) > 1, \quad \text{and} \nonumber\\
    \tilde M_{ba} &= M_{ba}/M_{aa} > \tilde \mu_b = \mu_b/\mu_a.
\end{align}
Since these low-dimensional parameters $M_{ab}$  and $M_{ba}$ are linear combinations of the
high-dimensional parameters $K_{ij}$, the conditions for bistability can be
decomposed into their relative contributions from the high-dimensional
interspecies feedbacks $K_{ij}$.}

\cchange{Specifically, consider the numerators of these inequalities, $M_{ab} =
\sum_{ij} \alpha_{ij}^{ab} K_{ij}$ and $M_{ba} = \sum_{ij} \alpha_{ij}^{ba}
K_{ij}$ (as in Eq.~(\ref{2Dto11D})). Then, the relative contributions to
$M_{ab}$ by each of the $\alpha_{ij}^{ab} K_{ij}$ terms
may be compared (and likewise for $M_{ba}$). When $\vec y_a$ corresponds to the CDI-susceptible
state and $\vec y_b$ corresponds to the CDI-resistant state,
the contributions to $M_{ab}$ are dominated by
the inhibition of Barnesiella on both Blautia and undefined genus of
Enterobacteriaceae (i.e., the contributions $\alpha_{9,1}^{ab} K_{9,1}$ and
$\alpha_{5,1}^{ab} K_{5,1}$). Contributions to $M_{ba}$ are dominated by
the inhibition of undefined genus of Enterobacteriaceae and Blautia on
unclassified Lachnospiraceae and Barnesiella (i.e., the contributions
$\alpha_{3,9}^{ba} K_{3,9}$, $\alpha_{3,5}^{ba} K_{3,5}$, $\alpha_{1,9}^{ba}
K_{1,9}$, and $\alpha_{1,5}^{ba} K_{1,5}$). Additional details about these
contributions are provided in the Supplementary Information. Thus, the
bistability between steady states $\vec y_a$ and $\vec y_b$ is largely driven
by feedbacks between a pair of species present in $\vec y_a$ (undefined genus of
Enterobacteriaceae and Blautia) and a pair of species present in $\vec y_b$
(unclassified Lachnospiraceae and Barnesiella).} 

\subsection{Bifurcation analysis guides interaction parameter changes that
modify steady-state outcomes in reduced 2D gLV systems}

Next, the bifurcation analysis of 2D gLV systems depicted in
Fig.~\ref{fig:bifur} indicates how to drive an initial condition 
$(0.5,0.5)$ towards
the target steady state $(1,0)$. This requires enlarging the basin of
attraction of the steady state $(1,0)$, which is equivalent to rotating the separatrix
counter-clockwise. The SSR-generated 2D system is bistable, and thus belongs to
the topological class in the upper-right quadrant 
of Fig.~\ref{fig:bifur}. Accordingly, the parameter $M_{ab}$ is decreased. When $M_{ab} =
M_{bb}\mu_a/\mu_b$, the \cchange{alternative} steady state at $(0,1)$ becomes unstable,
guaranteeing the initial condition $(0.5,0.5)$ will tend towards the
\cchange{target} state at $(1,0)$. 
However, to identify the minimal intervention that drives the system towards
the target state, we consider intermediate steps between the original value of
$M_{ab}$ and the bifurcation point $M_{bb}\mu_a/\mu_b$. 

Four incremental parameter changes
are plotted in Fig.~\ref{fig:Result_4_panel}C. On the fourth step,
the separatrix is sufficiently modified so that the initial condition tends
towards the target healthy steady state. The original 2D interaction matrix $M$, the
parameter change to $M_{ab}$, and the resulting interaction matrix
$M+\Delta M$ are visualized in Fig.~\ref{fig:matrix}E-G. The trajectory
plots in the bottom-left and the bottom-right corners of Fig.~\ref{fig:matrix}
illustrate the behavior of the 2D gLV system
parameterized by $M$ and $M+\Delta M$, respectively.
Therefore, SPARC can identify and modify interaction parameters to switch
the steady state behavior of this 2D model.

\subsection{SSR maps low-dimensional bifurcation behavior to the high-dimensional
system}

\begin{figure}
\includegraphics[width=1\columnwidth]{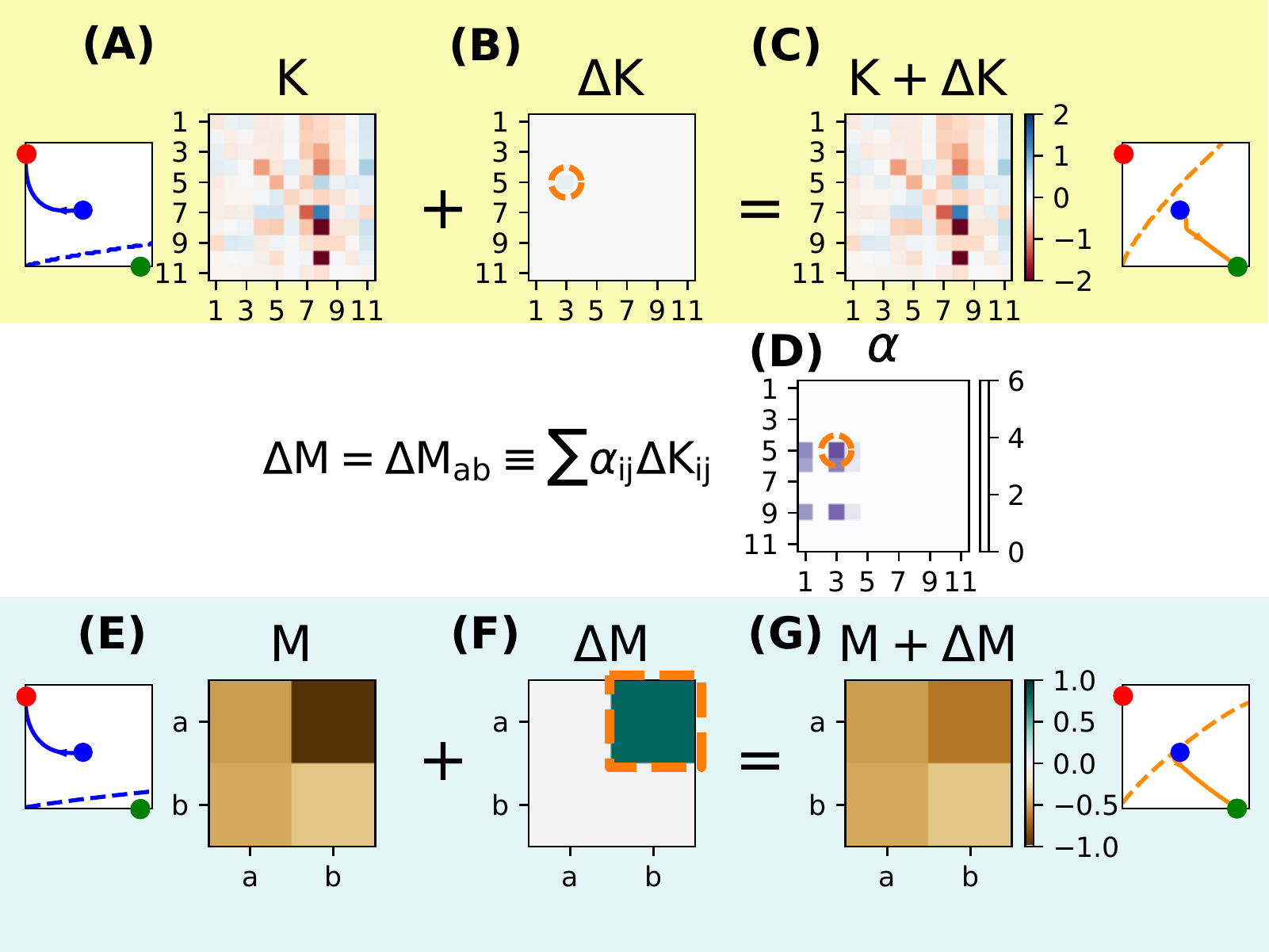}
    \caption{\label{fig:matrix}\textbf{Realization of the SSR-guided
    changes to gLV interaction parameters.} 
    As described in SPARC, the original high-dimensional
    interaction matrix $K$ (A), SSR-guided parameter change $\Delta K$ (B), and the
    resulting interaction matrix $K+\Delta K$ (C) are displayed. The
    steady-state reduced parameter matrix $M$ (E), bifurcation analysis guided
    parameter change $\Delta M$ (F), and the resulting 2D interaction matrix
    $M+\Delta M$ (G) are also displayed.
    The low-dimensional parameter change $\Delta M$, is related 
    to high-dimensional parameter changes through the SSR formulae
    Eq.~(\ref{SSR}).
    The $\alpha_{ij}$ coefficients represent the weights of the elements of the
    high-dimensional interaction matrix $K$ in the steady-state reduced interaction matrix $M$,
    as in Eq.~(\ref{2Dto11D})
    and these coefficients are visualized in panel (D). 
    To minimize the size of the high-dimensional parameter change,
    the interaction parameter $K_{ij}$ that corresponds to largest
    coefficients $\alpha_{ij}$ is chosen to be modified. 
    In this case, the coefficient $\alpha_{5,3}$ is
    the largest, which determines the choice of $\Delta K$.
    The phase space diagrams in each corner illustrate the trajectory of the
    initial condition $(0.5\vec y_a+0.5\vec y_b)$ or $(0.5,0.5)$, for each of the adjacent interaction matrices.
    }
\end{figure}

Finally, having determined the low-dimensional parameter modification that
alters the separatrix in the reduced 2D model (as shown in
Fig.~\ref{fig:Result_4_panel}C), corresponding high-dimensional parameters
that alter the system outcome in the original model can be identified.
Due to the degeneracy associated with mapping from the low-dimensional to
high-dimensional parameters, as is clear in the SSR formulae given by Eq.~(\ref{SSR}),
there are numerous 
modifications to the high-dimensional interaction matrix $K$ that correspond to the same
change in the 2D interaction matrix, as shown in
Fig.~\ref{fig:matrix}D.  In the CDI model, if the parameter change is confined
to only one 
element of $K$, there are a total of 121 choices. In order to make the
smallest change in the interaction matrix $K$, the coefficient $K_{ij}$
corresponding to the largest
$\alpha_{ij}$ value is chosen, as described in Eq.~(\ref{2Dto11D}). 
Specifically, the parameter change $\Delta K_{5,3} = 0.1744$ is used.

In Fig.~\ref{fig:matrix}D the magnitudes of the $\alpha_{ij}$ coefficients are
plotted, and the largest coefficient is highlighted with a dashed box. In the
bottom row of Fig.~\ref{fig:matrix}, the original $K$ matrix 
(panel A), the required modification $\Delta K$ corresponding to that
$\alpha_{ij}$ coefficient (panel B), and the resulting modified interaction matrix 
$K+\Delta K$ (panel C) are displayed. 
The trajectories in the upper-left and upper-right corners indicate the
behavior of the systems parameterized by $K$ and $K+\Delta K$, respectively.

Fig.~\ref{fig:Result_4_panel}D displays the results of
a representative 11D interaction matrix change $\Delta K$ that
drives the initial condition to the target state $\vec y_a$.
As in Fig.~\ref{fig:Result_4_panel}C, four incremental parameter
changes that each modify the separatrix are plotted. The largest of these four parameter
changes rotates the 11D separatrix counter-clockwise so that the initial condition
$(0.5\vec y_a + 0.5\vec y_b)$ tends towards the healthy steady state $\vec
y_a$.
Although small discrepancies exist between Fig.~\ref{fig:Result_4_panel}C and
Fig.~\ref{fig:Result_4_panel}D due to the SSR approximation, SPARC
successfully alters the steady-state outcome of a high-dimensional gLV
system by deliberately changing its interaction parameters. 

\setlength\belowcaptionskip{-2ex}

\subsection{\cchange{SPARC generates a finite-time intervention that drives a
disease-prone initial condition towards a healthy state in the CDI model}}

\begin{figure}
\includegraphics[width=1\columnwidth]{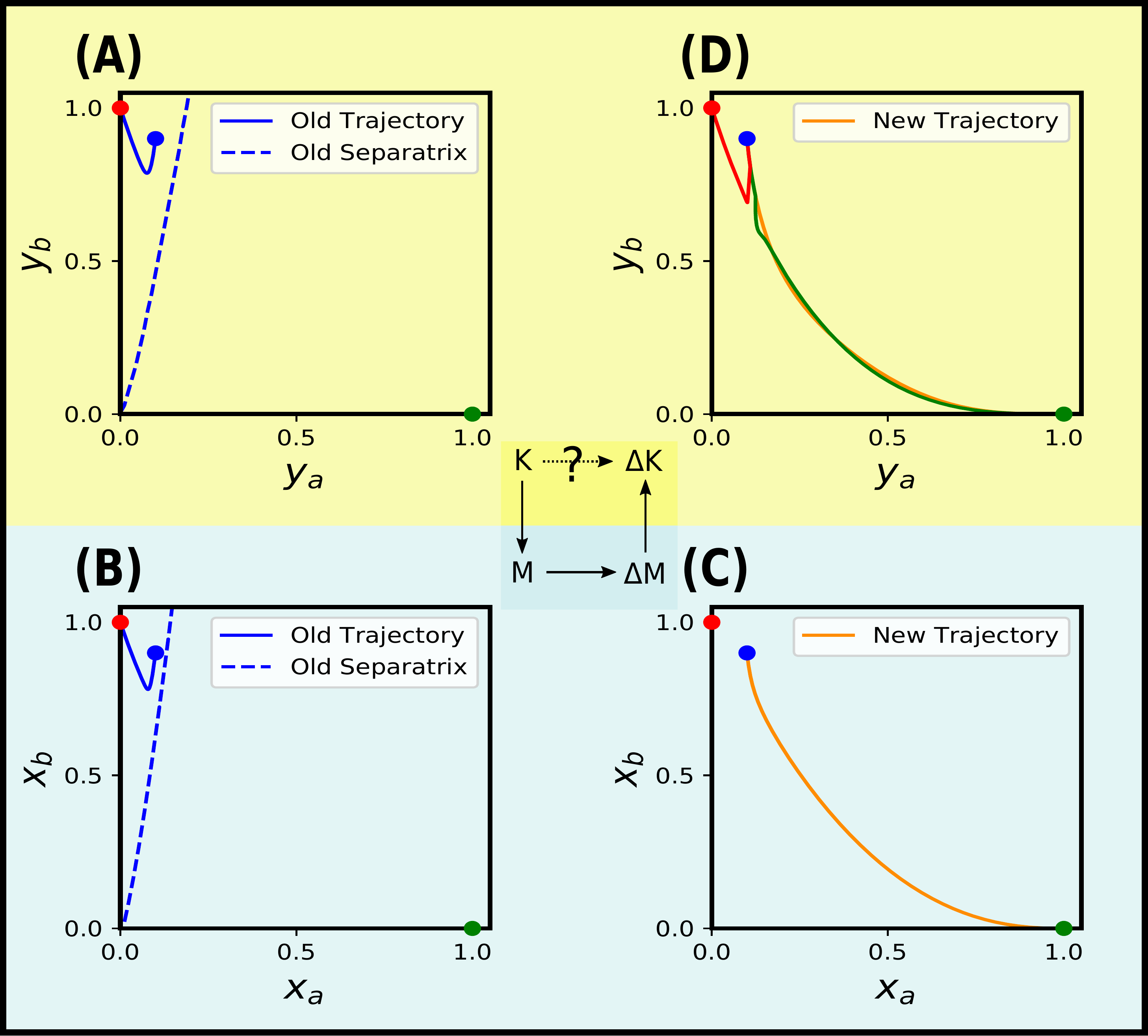}
    \caption{\label{fig:correct}\textbf{A realization of SPARC applied to the
    CDI model in the recovery scenario.} \cchange{Here, the target
    state $\vec y_a$ is the
    CDI-resistant state and the alternative state $\vec y_b$ is the CDI-susceptible
    state. \oran{The target and alternative steady states at  $(||\vec y_a||_2,0)$ and 
    $(0,||\vec y_b||_2)$ are rescaled in this plot to the points
    $(1,0)$ and $(0,1)$.}
(A) With the positions of the steady states switched, the 2D projection of
    the high-dimensional bistable region shown in Fig.~\ref{fig:Result_4_panel} is
    redrawn. Here the separatrix is close to the alternative state $\vec y_b$.
    The initial condition at $0.1\vec y_a + 0.9 \vec y_b$
    tends towards the alternative state $\vec y_b$. (B) The SSR formulae are applied to generate a
    2D approximate model. (C) After a parameter change $\Delta
    M_{ab}$, the steady state at $(0,1)$ becomes unstable and the initial
    condition now tends towards the target state. (D) The parameter
    change in the 2D model is associated with a parameter change in the original CDI model. 
    The yellow line plots the
    trajectory when the parameter modification is turned off after the system
    stabilizes, as in the case of Fig.~\ref{fig:Result_4_panel}. The red line
    shows the trajectory when the parameter change is turned off before the
    critical time, and the green line shows the trajectory when the parameter
    change is turned off after the critical time.
    }
        }
\end{figure}

\cchange{
    Next, we consider the} \oran{recovery} \cchange{ scenario in which the ``healthy"
CDI-resistant state is the target state $\vec y_a$ and the ``diseased" CDI-susceptible
state is the alternative state $\vec y_b$. The initial condition at $(0.1 \vec
y_a+ 0.9 \vec y_b)$ is chosen to demonstrate that SPARC can be effective even
when the initial condition is closer to the alternative state than to the
target state. As in the previous case, SPARC is applied
to change the steady-state outcome of this initial condition, which is shown in
Fig.~\ref{fig:correct}. For clarity, the shifted separatrices in
Fig.~\ref{fig:correct}C and D are not displayed. 
}

\cchange{
Without any parameter modification, 
the bistable region is exactly the reflection of the previous case,
as shown in Fig.~\ref{fig:correct}A and B. However, the parameter modification
generated by SPARC shifts the separatrix in the opposite direction. In this
case, the separatrix is already close to the alternative steady state at
$(0,1)$. The 2D parameter modification makes $\tilde M_{ab} <
1/\tilde \mu_b$, resulting in the steady state at $(0,1)$ becoming unstable, 
as shown in Fig.~\ref{fig:bifur} (top-right and top-left panels). Therefore,
although the initial condition is nearby the alternative steady state, after
modifying the low-dimensional parameters it
tends towards the target state at $(1,0)$.
}

\cchange{
The successful 2D parameter change is projected to the high-dimensional model.
Notably, the applied parameter change causes the steady state $\vec y_b$ in the
high-dimensional model to become unstable. Thus SPARC is capable of altering
the stability properties of high-dimensional steady states, which enables the
control of initial conditions even when they are located at or nearby an
alternative steady state.
}

\cchange{
Fig.~\ref{fig:correct}D also shows the effect of the duration of the parameter
modification. For SPARC to succeed, the parameter modification needs to be
active long enough for the microbial state to escape its original basin of
attraction. The red trajectory in 
Fig.~\ref{fig:correct} demonstrates that the system returns back to the alternative
steady state if the parameter change is applied for too short of a duration. The
green trajectory illustrates that the system will evolve towards the target
state as long as the parameter change is active beyond a critical duration.
This critical duration varies from case to case and was determined here
numerically by trial-and-error. 
The orange trajectory occurs when the parameter change is active until the system
stabilizes at the shifted steady state, as in Fig.~\ref{fig:Result_4_panel}.}
\setlength\belowcaptionskip{-2ex}

\subsection{SPARC successfully changes steady
state outcomes in synthetic gLV models}

\subsubsection{``Permuted'' synthetic models}

\begin{figure*}
\includegraphics[width=2\columnwidth]{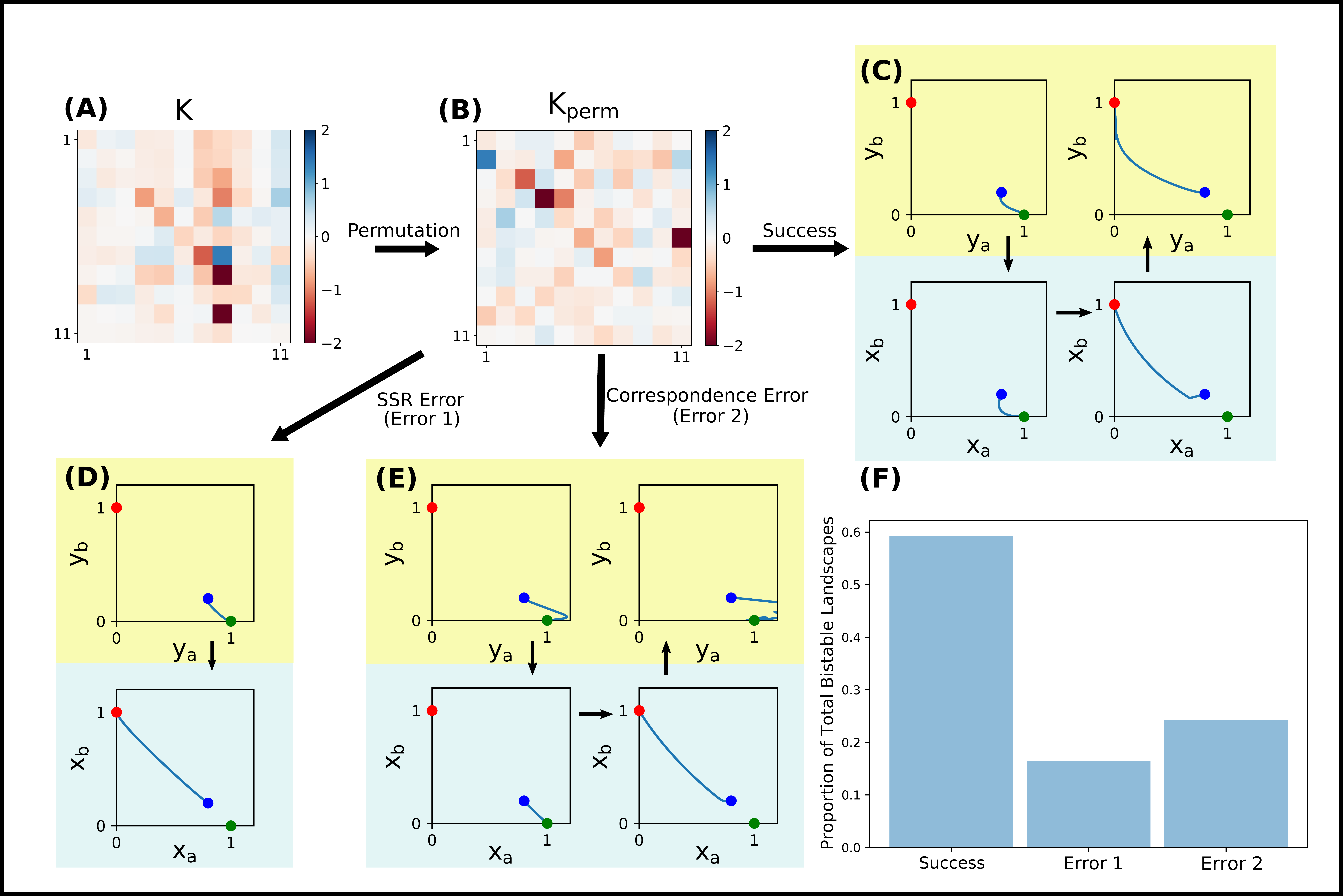}
    \caption{\label{fig:syn_stat}
    \cchange{\textbf{SPARC is effective at modifying steady-state outcomes in
    synthetic gLV models.}
    (A, B) The interaction matrix $K$ from the CDI model is randomly permuted to generate
    100 synthetic parameter sets. From these 100 synthetic gLV
    systems, 140 bistable regions are identified. 
    SPARC is applied to these synthetic models. (C)
    SPARC is considered successful if the parameter modification changes the
    trajectory of the initial condition so that it
    tends towards the target state (green), rather than the alternative
    state (red).
    (D, E) Two types of errors in SPARC are possible. 
    SPARC can fail during the steady-state reduction
    process if the outcome of the high-dimensional system does not agree
    with the steady-state outcome of the reduced system (SSR Error, panel D).
    It can also fail if the high-dimensional parameter change $\Delta K$ does not
    appropriately alter the steady-state outcome (Correspondence Error, panel E). 
    (F) SPARC successfully modifies 57\% (77/136) of the synthetically-generated
    bistable landscapes.
    These numbers represent a baseline error rate of
    SPARC that may be further improved through manual intervention.}}
\end{figure*}

To verify that SPARC is generalizable, it is applied to 100 
synthetic parameter sets generated by permuting the interaction parameters of the
CDI model. 
In these synthetic parameter sets, the growth rates $\rho_i$ 
are kept the same as in the CDI model. The
diagonal entries of the interaction matrix $K$ are all negative (as shown in
Fig.~\ref{fig:matrix}A), which is biologically reasonable since positive 
diagonal entries imply unphysical infinite
growth. 
To ensure the synthetic data
sets preserve this property, the diagonal and off-diagonal entries of the
$K$ matrix are permuted independently. All 100 parameter sets are generated in this way.
This permutation process is demonstrated in Fig.~\ref{fig:syn_stat}A and B.

In the next step, bistable regions for each synthetic system must be
identified in order for SPARC to be applicable.
\cchange{Steady state analysis shows that, for a randomly permuted parameter
set, stable steady states are small in number. From 100 permuted gLV
parameter sets, there are on average $0.8$ completely stable steady states 
and $5.3$ steady states with at most one unstable direction (i.e.,
steady states whose Jacobian matrices have at most one non-negative eigenvalue)
per parameter set.}

\cchange{To ensure there are enough steady states to form bistable
landscapes,} we compute all $2^N$ steady states of each synthetic
parameter set, then identify all steady states whose Jacobian has 0 or 1
positive eigenvalues in each parameter set, and use numerical simulations to test whether 
each steady state pair forms a bistable landscape.
Specifically, for a steady state pair $\vec y_a$ and $\vec y_b$, trajectories
with initial conditions $(0.95\vec y_a + 0.05 \vec y_b)$ and $(0.05 \vec y_a+0.95 \vec y_b)$
are simulated to test whether they tend towards their nearest steady
state. In addition, if initial conditions at $(0.8\vec y_a+ 0.2 \vec y_b)$
or $(0.2\vec y_a + 0.8 \vec y_b)$ tend towards some other third steady
state, the steady state pair is excluded. 
Out of the 100 synthetic parameter sets, a total of 136 bistable landscapes
were identified. 

In this context, SPARC is considered successful if it identifies
high-dimensional interaction
parameter changes that alter the steady-state outcome in a
bistable system, \cchange{as in Fig.~\ref{fig:syn_stat}C}. This success relies on the
correspondence between the 11D and 2D landscapes generated by SSR, the
bifurcation analysis of the 2D system, and the correspondence between 2D and
11D parameters governed by the SSR formulae. Therefore, if an initial condition
in both the unperturbed 11D and 2D models tends towards the same steady state, 
and the same initial condition in both the perturbed
11D and 2D models tend towards the other steady state in the bistable landscape, 
SPARC is considered successful. 

\setlength\belowcaptionskip{-2ex}

To examine the fidelity of SPARC on synthetic parameter sets, it is applied to an
ensemble of synthetically generated models. 
The two steady states of the synthetic bistable system are arbitrarily labeled as $\vec
y_a$ and $\vec y_b$.
In Fig.~\ref{fig:Result_4_panel} the initial condition was located at $(0.5\vec
y_a +0.5 \vec y_b)$, but for these synthetic parameter sets the 
initial condition is located at $(0.2 \vec y_a + 0.8 \vec y_b)$. 
Since SSR is more accurate near the two steady states, this choice of
initial condition improves the success rate of SPARC.
SPARC can fail at two steps, corresponding to the arrows A to B
and C to D in the schematic Fig.~\ref{fig:schematic}.
The first type of error occurs when SSR fails to preserve the steady
state behavior of the gLV model; \cchange{this error is demonstrated in
Fig.~\ref{fig:syn_stat}D, where the high-dimensional initial condition tends
towards steady state $\vec y_b$ but the initial condition of the SSR-reduced
model tends towards steady state $\vec x_a$}.
The second type of error occurs when associating the low-dimensional
parameter change with a high-dimensional parameter change; \cchange{this error
is demonstrated in Fig.~\ref{fig:syn_stat}E, where the modified low-dimensional
trajectory correctly tends towards steady state $\vec x_a$, but its
corresponding high-dimensional trajectory erroneously tends towards steady
state $\vec y_b$.} 
Since the choice of a high-dimensional parameter change is degenerate,
modifications to four interaction parameters $K_{ij}$ corresponding to the four largest
$\alpha_{ij}$ coefficients are tested. 
Small changes in the interaction matrix $K$ will
slightly change the location of the steady states, so this perturbation is
turned off after the system has relaxed to the shifted steady states to allow
the system to return to its original steady states.

Out of the 136 bistable landscapes generated from 100
synthetic parameter sets,
SPARC successfully identified parameter modifications that led
to the targeted transition between steady-state outcomes 57\% (77/136) of the time. 
\cchange{Details about specific
errors rates occurred are provided in Fig.~\ref{fig:syn_stat}F}: 17\% (23/136) occurred during
the SSR compression step, and 26\% (36/136)
occurred during the mapping from the 2D parameters to high-dimensional
parameters.
\oran{
Manual intervention (e.g., trying different sizes of the prescribed
parameter change) can improve this success rate.
Therefore, SPARC is effective at altering steady-state
behavior in generic gLV systems. 
}

\subsubsection{``Noisy'' synthetic models}

\cchange{Finally, since inferring parameter values in gLV systems is an intrinsically
noisy procedure, it is valuable to understand whether parameter changes
generated by SPARC are robust to noise in the fitted parameters.  We consider
the ``infection'' scenario in which the CDI-susceptible state is the target
state $\vec y_a$ and the CDI-resistant state is the alternative state $\vec y_b$.
\oran{Implementing the parameter change prescribed by SPARC ($\Delta K_{5,3}=0.1744$)
successfully drives an initial condition at $(0.5 \vec{y}_a + 0.5\vec y_b)$
towards the target state $\vec y_a$.} 
To test the robustness of the SPARC method, this
parameter change is applied to synthetic ``noisy'' systems that are
generated by independently scaling each interaction parameter $K_{ij}$ by a number
randomly drawn from a uniform distribution between $1-\beta$ and $1+\beta$. This parameter $\beta$
is called the ``noise.''}
\oran{
In the Supplementary Information the following analysis is also
performed using a parameter change 20\% larger than the original one ($\Delta
K_{5,3}=0.2092$). This increment compensates for the deviation between the
original and the SSR-generated separatrices.
}

\cchange{The steady states of these synthetic systems
are functions of the interaction parameters, and therefore differ from the
steady states of the original CDI model. In gLV systems
the presence/absence combination of species uniquely identifies a steady state,
so it is straightforward to identify the two steady states in these noisy
systems, called $\tilde{ \vec y}_a$ and $\tilde{\vec y}_b$, that correspond to
the target and
alternative steady states $\vec y_a$ and $\vec y_b$ of the original CDI model.
Many of these newly-generated steady states are biologically unreasonable: for
a noise of $\beta=0.025$, nearly half of the noisy steady states
$\tilde{\vec y}_a$ contain negative entries. Additional details
regarding the deviations of the noisy steady states as a function of the
noise $\beta$ are provided in the Supplementary Information.}

\cchange{
We only consider noisy synthetic systems (i) that do not contain any negative
entries in the steady states $\tilde{ \vec y}_a$ and $\tilde{\vec y}_b$, and
(ii) in which an initial condition at $(0.5 \vec{y}_a + 0.5\vec y_b)$ flows
towards the alternative state $\tilde{\vec y}_b$ in the absence of any
intervention (note that the initial condition is based on steady states of the
original CDI model). Then, the parameter change $\Delta K_{5,3} = 0.1744$ is
applied to the noisy models for an initial condition $(0.5 \vec{y}_a + 0.5\vec
y_b)$; if the system flows towards the target state $\tilde{\vec y}_a$ the
parameter change is considered successful, and if it does not it is considered
an error.  The error rate of SPARC as a function of the noise $\beta$ is
plotted in the Supplementary Information. For each noise value, 1000 synthetic
systems are created to generate statistics for the error rates.  
\oran{
Using the original parameter change $\Delta K_{5,3} = 0.1744$
works well for very small noise values ($\beta < 0.005$), rapidly increases to
an error rate of 40\% with a noise of $\beta = 0.02$, and eventually approaches
80\% for a noise of $\beta = 0.5$.
}
With the incremented parameter change of $\Delta K_{5,3}=0.2092$, 
the CDI model works nearly perfectly for
synthetic systems with small noise values ($\beta < 0.02$). \oran{Then, as the
noise increases} the error rate worsens: a noise of $\beta = 0.1$
corresponds to a 30\% error rate, and a noise of $\beta = 0.5$ approaches an
80\% error rate. These analyses indicate that interventions generated by SPARC
are effective for gLV systems whose parameters are known precisely, but are
less effective when parameters are relatively unconstrained. Taken another way,
these results place a limit on the required accuracy of parameter estimation,
beyond which point two measured systems will differ enough in their parameter
values that they diverge in their behavior. }

\section{Discussion}
\subsection{SPARC is efficient and flexible}

SPARC generates a 2D gLV model to guide high-dimensional
parameter modifications that alter the system outcome. 
Without such a guide, this parameter change must be instead selected
through trial-and-error. A study about T-cell cancer networks used this
exhaustive trial-and-error method to 
find parameter perturbations that drive the system between attractors, but it
was computationally expensive to search their parameter space \cite{Wang2016}.  
In gLV systems, 
the number of computations needed for this trial-and-error method 
grows as $\mathcal{O}(N^2)$, where $N$
is the number of species in the gLV model. As the number of species $N$ becomes
large, the exhaustive method becomes
computationally intractable.

Rather than exploring the $N^2$-dimensional parameter space of $K$, 
SPARC allows exploration of a 2-dimensional subspace of $M$
associated with the bistable dynamics of interest of the high-dimensional model.
In the SSR-generated 2D model,
parameter modifications are analytically tractable using bifurcation analysis,
which determines the sign of the parameter change according to the direction
of the required separatrix shift.
After the 2D model parameter change is determined, SSR formulae provide a
direct correspondence between the 2D and the high-dimensional parameter
modifications that produce the same steady-state outcome.
\cchange{For example, since bistability is well-defined in the 2D gLV system,
SSR reveals the interspecies feedbacks most responsible for bistability in the
high-dimensional system.}

\oran{
    Furthermore, SPARC is flexible enough to drive
    the dynamical system bidirectionally between steady states, as demonstrated 
    in the infection and recovery scenarios. When one steady state
    in the bistable region is
    desirable, as in the clinically-motivated recovery scenario considered here,
    SPARC identifies both which parameter changes to avoid and which to
    perform in order to achieve the target outcome. Both types of parameter
    changes are informative when trying to prevent the system from tending
    toward an undesirable steady state.}

\cchange{Finally, we note that the applicability of the parameter changes recommended by SPARC is
sensitive to the accuracy of the fitted gLV interaction parameters.
For example, in the CDI system (as demonstrated by the noisy synthetic
models), the SPARC parameter change becomes less effective as the noise in the interaction parameters increases. 
This analysis quantifies the tolerable level of uncertainty 
in fitted interaction parameters
before they result in fundamentally different classes of model behavior.}

\subsection{Perturbing ecological interactions indirectly controls steady-state outcomes}

Direct control methods modify the 
steady-state outcome of the gut microbiome by changing the state of the
microbial system while retaining the same dynamical landscape. 
Implementations of this direct control method include bacteriotherapies such as
Fecal Microbial Transplantation (FMT), which has been shown to be an effective
treatment for 
\textit{Clostridioides difficile} infection (CDI). FMT introduces a foreign
microbial transplant that alters a host's microbiome composition, thereby
ameliorating symptoms of CDI \cite{FMT}.
As realized in the gLV model, this amounts to an instantaneous shift in the
microbial composition that moves the microbial state from one basin of attraction
to another.

In contrast to this direct control method, SPARC indirectly controls
the steady-state outcomes of a high-dimensional 
gLV model by modifying its dynamical landscape. 
Instead of adding foreign microbes, SPARC perturbs the
interaction parameters of the gLV model, which we interpret as changing the
environment in which the microbes live.
Fig.~\ref{fig:Result_4_panel} illustrates how this
parameter-altering control method changes the steady-state outcome of a
simulated gut microbial system. 

SPARC could be applied to other ecological
systems in order to attain a target community structure. In marine ecosystems, the
target community structures may correspond to ecological states without harmful
algal blooms or invasive fish species. In these cases, environmental factors 
such as the abundance 
of chemical fertilizers or pesticides, the pH, and the velocity of
stream flows influence the state of the ecosystem
\cite{rotenone,Algalbloom}. Previously, algal blooms and population dynamics of
invasive fish species have been modeled with gLV systems
\cite{Scotti2015,Whipple2000}. 
Therefore, SPARC could provide a systematic
framework that guides environmental interventions to remove harmful algae or
invasive fish species.

\cchange{
SPARC identifies a single entry in a high-dimensional interaction matrix that 
can be altered to change the system
behavior. However, it might not be possible in practice to identify
environmental factors that, when modified, change only
one entry of the interaction matrix. Importantly, the parameter entry 
generated by SPARC is not
unique, as shown in Eq.~(\ref{2Dto11D}). As a result, it is possible to find a
linear combination of changes in the environmental factors that maximize the parameter
changes in the most effective entries (i.e., entries with the largest $\alpha_{ij}$
values) and minimize other changes}, \oran{especially the most effective
entries in the opposite direction}. \cchange{This more complex
parameter change can then be simulated to assess its effectiveness.
}

\subsection{SPARC provides a lens for understanding the effect of the 
environment on microbial composition }

Having demonstrated the effectiveness of SPARC \textit{in
silico}, it would be valuable to verify this method in an experimental model system of
the microbiome.
SPARC relies on changing interactions between microbial species in the gut microbiome, 
which could be achieved by deliberately changing environmental factors in a
controlled experimental setting. 
Therefore, any realization of this method would require an experimental
microbiome model of limited microbial diversity that allows
the manipulation of oxygen levels, nutrient availability, or other
factors. One such experimental model might be the intestine-on-a-chip system,
which simulates the human gut
microbiome in a manipulable \textit{in vitro} environment 
\cite{oxygen,DeFilippo2010,intestine-on-a-chip}.  
By fitting gLV models to time-series data from the intestine-on-a-chip, 
it may be possible to isolate the effect of environmental perturbations and identify the
corresponding interaction matrix change $\Delta K$ underlying SPARC.

In real microbial systems, changes in environmental factors potentially affect
the interactions between many species, thus changing multiple interaction
parameters at a time.
For example, Lin \textit{et al.}~found that four dominant bacterial genera with carbon
assimilation pathways gain ecological advantages when there is a lack of 
dissolved carbon in the environment
\cite{Lin2015}. Therefore, environmental changes such as the removal of
dissolved carbon will alter the effective microbe-microbe 
interactions between these species.
In cases such as these, SPARC could systematically specify how
environmental changes alter the dynamical landscape. 

In future applications, the
environmental degrees of freedom will be as myriad as diet, designer
probiotics, or designer prebiotics. The combinatorial complexity of these 
contributions will require a systematic framework, such as SPARC, in order to
understand how to drive the system towards a target state.
Once environmental interventions are associated with changes in species-species
interaction parameters in gLV models, 
SPARC could help predict how environmental changes affect gut microbiome compositions.

\section{Conclusion}
SPARC controls the steady-state outcome of bistable regions in gLV
systems by altering ecological interaction parameters. 
This method circumvents the computational task of performing numerical
trials to exhuastively search a
high-dimensional parameter space. Instead, SPARC uses a recently-developed
dimensionality-reduction technique to reduce the problem to searching a
2-dimensional parameter subspace. Consequently, we are able to efficiently 
and systematically identify a minimal parameter change that 
results in desired system behavior.

SPARC provides a novel alternative to canonical control methods that
modify the system state directly. 
SPARC instead focuses on how environmental factors and microbial
interactions dictate microbial dynamics. Eventually, indirect and direct
methods could be used in conjunction to provide a comprehensive framework for
the control of ecological systems. 





\begin{acknowledgments}
This material was based upon work supported by the
National Science Foundation Graduate Research Fellowship
Program under Grant No.~1650114. Any opinions, findings,
and conclusions or recommendations expressed in this material 
are those of the author(s) and do not necessarily reflect
the views of the National Science Foundation. This work was
also supported by the David and Lucile Packard Foundation
and the Institute for Collaborative Biotechnologies through
Contract No. W911NF-09-D-0001 from the U.S. Army Research Office. 
The funders had no role in study design, data
collection and analysis, decision to publish, or preparation of
the manuscript.
\end{acknowledgments}



\raggedright

%


\end{document}